\documentclass[letterpaper, 10 pt, conference]{ieeeconf}
\IEEEoverridecommandlockouts
\usepackage[utf8]{inputenc}
\usepackage{courier}
\usepackage{cite}
\usepackage{graphicx}
\usepackage[cmex10]{amsmath}
\usepackage{array}
\usepackage{color}
\usepackage{amssymb}
\usepackage{amsfonts}
\usepackage[short]{optidef}
\usepackage{balance}
\usepackage{subcaption}
\usepackage{float}
\usepackage{subfiles}
\usepackage{svg}
\usepackage{comment}
\usepackage{multirow}
\usepackage{multicol}
\usepackage{arydshln}
\usepackage[font=small,skip=0pt]{caption}

\newtheorem{condition}{Condition}
\newtheorem{theorem}{Theorem}

\newcommand{\stox}{{x}}

\newcommand{\stou}{{u}}

\newcommand{\stoz}{{z}}
\newcommand{\tx}[1]{{#1}}

\let\svthefootnote\thefootnote
\newcommand{\thx}{\let\thefootnote\relax\footnotetext{\hspace*{0mm} This work is funded by the European Union (Horizon Europe, ERC, COMPLETE, 101075836). Views and opinions expressed are however those of the author(s) only and do not necessarily reflect those of the European Union or the European Research Council Executive Agency. Neither the European Union nor the granting authority can be held responsible for them.}
\let\thefootnote\svthefootnote}

\title{\LARGE \bf Learning-based model augmentation with LFRs}

\author{Jan H. Hoekstra$^1$, Chris Verhoek$^1$, Roland Tóth$^{1,2}$, and Maarten Schoukens$^1$ \\
\small $^1$ Control Systems Group, Eindhoven University of Technology, The Netherlands \\
\small $^2$ Systems and Control Laboratory, Institute for Computer Science and Control, Budapest, Hungary \\
\small j.h.hoekstra@tue.nl, c.verhoek@tue.nl, r.toth@tue.nl, m.schoukens@tue.nl}

\begin{document}

\maketitle \thx

\begin{abstract}
    \emph{Nonlinear system identification}~(NL-SI) has proven to be effective in obtaining accurate models for highly complex systems. In particular, recent encoder-based methods for \emph{artificial neural networks state-space} (ANN-SS) models have achieved state-of-the-art performance on various benchmarks, while offering consistency and computational efficiency. Inclusion of prior knowledge of the system can be exploited to increase (i) estimation speed, (ii) accuracy, and (iii) interpretability of the resulting models. This paper proposes an encoder-based model augmentation method that incorporates prior knowledge from \emph{first-principles} (FP) models. We introduce a novel \emph{linear-fractional-representation} (LFR) model structure that allows for the unified representation of various augmentation structures including the ones that are commonly used in the literature, and an identification algorithm for estimating the proposed structure together with appropriate initialization methods. The performance and generalization capabilities of the proposed method are demonstrated in a hardening mass-spring-damper simulation.
\end{abstract}
\begin{keywords}
    Nonlinear System Identification, Model Augmentation, Physics-based Learning, Linear Fractional Representation
\end{keywords}

 \label{sec:Abstract}

\vspace*{-1pt}
\section{Introduction} \label{sec:Introduction}
As systems used in control become increasingly complex and performance requirements increase, the demand for accurate nonlinear models that capture complicated behaviors is rapidly growing. Obtaining these models through \emph{first-principles} (FP) based methods is becoming too costly and often infeasible due to the surge of system complexity and difficult to model phenomena for which no reliable FP descriptions exist. To overcome these issues, \emph{nonlinear system identification} (NL-SI) methods can be used to estimate models starting from measurements of the system\cite{Schoukens2019}. In particular, the estimation of black-box models combined with \emph{artificial neural networks} (ANNs) has been able to achieve unprecedented accuracy. In control, ANN \emph{state-space} (SS) models have been used to cope with high-order systems and capture complex nonlinear dynamic effects\cite{suykens1995, Schoukens2021}. Recent encoder-based methods for ANN-SS models~\cite{gerben2022,Forgione2021TruncSimulation, Beintema2022} have shown state-of-the-art performance on a wide range of benchmark problems, while providing consistency guarantees and computational efficiency.

Although the identified models obtained with black-box methods may be accurate, they are not based on physical laws and are difficult to interpret. This forms a substantial roadblock to the use of these models in industry, where interpretable models are preferred for the design process\cite{ljung2010perspectives, Panel2021}. Furthermore, black-box methods do not consider the prior knowledge available of the system, such as (approximate) FP models that have been created in the past. As a result, black-box methods will discover already known dynamics, which is an apparent waste of prior knowledge and modeling resources and often results in excessive amount of estimation time.

Physics-based learning methods can include prior information of the system in the NL-SI algorithm in a variety of ways\cite{willard2020,Daw2022, psichogios1992hybrid, pathak2018hybrid}, resulting in (i) faster estimation and (ii) better accuracy on unseen data than black-box methods. Model augmentation is a promising approach in which prior information is included by combining an FP-based model with flexible function approximators in a variety of different model augmentation structures, such as parallel\cite{sun2020comprehensive} and series\cite{gotte2022composed, Groote2022, Shah2022} interconnections. Such augmentation structures can result in (iii) more interpretable models.

The model augmentation methods mentioned above produce models with benefits (i)-(iii) compared to ANN-SS identification methods; however, they often require full-state measurements\cite{sun2020comprehensive,gotte2022composed, Groote2022, Shah2022} or an estimation of the states\cite{bohlin2006practical}. In practice, often only input-output data is available. Secondly, existing model augmentation structures only consider a limited number of interconnections between the learning components and the baseline model. Such interconnections might be necessary for accurate augmentation of the state transition function or adding missing dynamic (states) to the baseline model. A more flexible model augmentation structure and algorithms capable of reliable model learning with it would be required to consider such interconnections. 

The encoder-based ANN-SS methods can handle input-output data efficiently and have shown state-of-the-art results in deep learning-based system identification\cite{gerben2022,Forgione2021TruncSimulation, Beintema2022}. Hence, a combination of model augmentation and encoder-based methods seems promising. Despite this, the extension of encoder-based methods to model augmentation scenarios has been limited to a case study with promising results\cite{Retzler2024}.

We propose a model-augmentation-based identification method building on recent results in NL-SI in terms of \emph{Sub-Space Encoder Network} (SUBNET) \cite{gerben2022,Beintema2021,Beintema2022}. Our contribution in this paper is the adaption of SUBNET to the model-augmentation identification problem. To achieve this, we propose a novel \emph{linear-fractional-representation}~(LFR)\cite{redheffer1960certain} based model augmentation structure that flexibly combines the baseline model with a black-box learning component. In the LFR model structure, an interconnection matrix governs the signals between the baseline model and the augmentation. By shaping this interconnection matrix, we can represent a wide range of model augmentation structures proposed in the literature~\cite{gotte2022composed, sohlberg2008grey, thompson1994modeling, Daw2022,Bolderman2024, Schoukens2020} and a wide variety of structures beyond them, resulting in an unifying model structure represented in LFR form. The LFR model structure also allows for the addition of additional dynamical components, in terms of states of the learning component, beyond the baseline model dynamics. Furthermore, we also consider joint identification of the baseline model parameters and the learning component parameters and handle the problem of overparameterization through regularization\cite{Bolderman2024}. The contributions are summarized as follows:
\begin{itemize}
    \item We propose a LFR-based model augmentation structure unifying existing model augmentation structures;
    \item We extend the LFR form for dynamic model augmentation to better capture missing dynamics;
    \item We provide a system identification algorithm capable of joint estimating the proposed LFR-based model augmentation structure with ANN parametrization of the learning component.
\end{itemize}

In the remainder of the paper, first the NL-SI problem is introduced with baseline models in Section~\ref{sec:Problem}, followed by the introduction of the LFR-based model augmentation structure, dynamic augmentation, and system identification algorithm in Section~\ref{sec:Method}. A hardening \emph{mass-spring-damper} (MSD) simulation example is used to demonstrate the performance of the proposed identification method and compare it to black-box ANN-SS methods in Section~\ref{sec:Example}. Conclusions are given in Section~\ref{sec:Conclusion}.

\section{Identification Problem} \label{sec:Problem}
We consider the data-generating system given by the \emph{discrete-time} (DT) nonlinear representation
\vspace*{-1mm}
\begin{subequations}\label{eq:nl_dyn}
	\begin{align}
		x_{k+1}&=f(x_k,u_k),\\
		y_k&=h(x_k) +e_k,
	\end{align} 
\end{subequations} 
where $x_k \in \mathbb{R}^{n_\mathrm{x}}$ is the state, $u_k \in \mathbb{R}^{n_\mathrm{u}}$ is the input, $y_k \in \mathbb{R}^{n_\mathrm{y}}$ is the output signal of the system at time moment $k\in\mathbb{Z}$ with $e_k$ an i.i.d., possibly colored, noise process with finite variance representing measurement noise. The state-transition function $f: \mathbb{R}^{n_\mathrm{x}} \times \mathbb{R}^{n_\mathrm{u}} \rightarrow \mathbb{R}^{n_\mathrm{x}}$ and the output function $h: \mathbb{R}^{n_\mathrm{x}}\rightarrow \mathbb{R}^{n_\mathrm{y}}$ are considered to be bounded.

A baseline model for system \eqref{eq:nl_dyn} can, for example, be obtained by FP methods. The baseline model is given as
\vspace*{-5mm}
\begin{subequations}\label{eq:FP_model}
    \begin{align}
        \tilde{x}_{k+1} &=f_\text{base}\left(\theta_\text{base}, \tilde{x}_k, u_k\right), \\
        \tilde{y}_k &=h_\text{base}\left(\theta_\text{base}, \tilde{x}_k, u_k\right),
    \end{align}
\end{subequations}
where $\tilde{x}_k\in \mathbb{R}^{n_\mathrm{\tilde x}}$ is the baseline model state, $\tilde{y}_k \in \mathbb{R}^{n_\mathrm{y}}$ the baseline model output, and $f_\text{base}$ and $h_\text{base}$ are the baseline model functions for state transition and output, respectively, with the parameters $\theta_\text{base}$. These functions are jointly denoted as $\phi_\text{base}$. These baseline models can be good approximations of the system \eqref{eq:nl_dyn}. However, as stated in the introduction, obtaining accurate baseline models is costly and often infeasible. We, thus, utilize an easier to obtain approximate baseline models with, for example, incomplete dynamics and approximate parameters $\theta_\text{base}$. We then augment these baseline models with learning components in a model augmentation method.

In model augmentation, our aim is to represent \eqref{eq:nl_dyn} by a model that combines the baseline model \eqref{eq:FP_model} with learning components. Using a data sequence $\mathcal{D}_N=\left\{\left(y_k, u_k\right)\right\}_{k=1}^N$ generated by~\eqref{eq:nl_dyn}, the aim is to identify a DT model augmentation structure of the form
\vspace*{-2mm}
\begin{subequations}\label{eq:ss_model_structure}
    \begin{align}
        \hat{x}_{k+1} &= f_\text{base} \star f_\text{aug} \left(\theta_\text{base}, \theta_\text{aug}, \hat{x}_k, u_k\right), \\
        \hat{y}_k &= h_\text{base} \star h_\text{aug} \left(\theta_\text{base}, \theta_\text{aug}, \hat{x}_k, u_k\right),
    \end{align}
\end{subequations}
where $\hat{x}_ k\in \mathbb{R}^{n_{\hat{x}}}$ is the model state, $\hat{y}_k \in \mathbb{R}^{n_\mathrm{y}}$ is the model output, and $\star$ is an interconnection between the baseline model and the learning components $f_\text{aug}$ and $h_\text{aug}$ with the parameters $\theta_\text{aug}$. In the literature a variety of different realizations of this general model augmentation structure are presented, such as static parallel\cite{sun2020comprehensive} and static series~\cite{gotte2022composed, Groote2022, Shah2022} structures, as shown in Table~\ref{tab:augmentation_structures}. We consider these model structures as static augmentations, as they do not add additional states beyond the baseline model to capture dynamics missing from the baseline model. Dynamic augmentation structures such as \cite{bohlin2006practical} have also been considered. In this case, the model state $\hat x$ is $\left[\begin{array}{cc} \tilde x^\top & \bar x^\top \end{array}\right]^\top$, where $\bar x$ is the additional state added for the dynamic learning component.

These model augmentation structures found in the literature are limited in the interconnections considered between the baseline model and the learning components. For example, we consider novel dynamic augmentations in which the learning components take the model state $\hat x$ as input, and their output affects the entire model state $\hat x$ as shown in Table~\ref{tab:augmentation_structures}. This dynamic augmentation, as well as other augmentation structures, could be represented by a more flexible model augmentation structure representation of \eqref{eq:ss_model_structure}.

\begin{table}[t]
    \centering
    \caption{Classes of model augmentation structures.}
    \begin{tabular}{c|c|c}
        \hline
         & static parallel & static series \\
        \hline
        $\tilde x_{k+1}$ & $f_\text{base}\left(\tilde x_k, u_k\right) + f_\text{aug}\left(\tilde x_k, u_k\right)$ & $f_\text{aug}\left(f_\text{base}\left(\tilde x_k, u_k\right)\right)$ \\ 
        $\bar x_{k+1}$ & - & - \\

        \hline
         & dynamic parallel & dynamic series \\
        \hline
        $\tilde x_{k+1}$ & $f_\text{base}\left(\tilde x_k, u_k\right) + f_\text{aug}\left(\tilde x_k, \bar x_k, u_k\right)$ & $f_\text{aug}\left(f_\text{base}\left(\tilde x_k, u_k\right)\right)$ \\ 
        $\bar x_{k+1}$ & $g_\text{aug}\left(\tilde x_k, \bar x_k, u_k\right)$ & $g_\text{aug}\left(\tilde x_k, \bar x_k, u_k\right)$\\
        
        \hline 
    \end{tabular}
    \vspace{-10pt}
    \label{tab:augmentation_structures}
\end{table}

The proposed model augmentation structure should be able to represent the model structures in Table~\ref{tab:augmentation_structures} and a wide variety of model augmentation arrangements beyond them. Such a flexible model structure will require an algorithm capable of reliable joint estimation of the baseline $\theta_\text{base}$ and augmentation parameters $\theta_\text{aug}$. To address these limitations, we propose the adaptation of SUBNET\cite{gerben2022,Beintema2021,Beintema2022} to the model augmentation setting and a novel LFR-based model augmentation structure. We consider ANNs\footnote{Consider that each hidden layer is composed of $m$ activation functions $\rho:\mathbb{R} \rightarrow \mathbb{R}$ in the form of $\stoz_{i,j} = \rho(\sum_{l=1}^{m_{i-1}}\theta_{\mathrm{w},i,j,l} \stoz_{i-1,l}+ \theta_{\mathrm{b},i,j})$ where $\stoz_i=\mathrm{col}(z_{i,1},\ldots,z_{i,{m_i}})$  is the latent variable representing the output of layer $1\leq i\leq q$. Here, $\mathrm{col}(\centerdot)$ denotes the composition of a column vector. For a $f_\tx{\theta}$ with $q$ hidden layers and linear input and output layers, this means $f_\tx{\theta}(\hat{\stox}_\tx{k}, \stou_\tx{k})= \theta_{\mathrm{w},q+1} \stoz_q(k) + \theta_{\mathrm{b},q+1}$ and $\stoz_{0}(k)=\mathrm{col}(\hat{x}_\tx{k}, \stou_\tx{k})$.} as learning components.

\section{Method} \label{sec:Method}
\subsection{LFR-based augmentation structure}
A key property of LFR representations is that the elementary operations of summation, multiplication, and inversion allow manipulations with LFRs, and an LFR of an LFR is also an LFR\cite{redheffer1960certain}. The flexibility of this representation has made it popular in the field of robust control \cite{Zhou1996, junnarkar2022synthesis} and linear parameter-varying-control \cite{Toth2010, Schoukens2018}. Furthermore, an LFR can also include nonlinear components in the interconnections\cite{veenman2016robust}, which has made LFRs useful for black-box nonlinear system representations\cite{Schoukens2020, shakib2022computationally} and implicit learning\cite{el2021implicit}. Recent results have also shown that stability properties can be enforced on these black-box LFRs at the cost of some model accuracy\cite{revay2020convex, revay2021recurrent, frank2022robust}. In this work, we will use LFRs to include a baseline model of the system in a rather flexible arrangement w.r.t. the learning component.

\begin{figure}[t]
    \centering
    \includegraphics[width=0.34\textwidth]{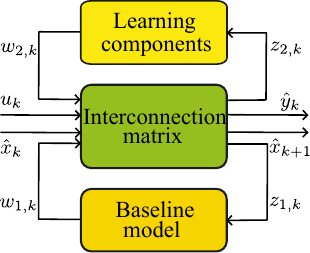}
    \vspace{5pt}
    \caption{LFR-based model augmentation of a baseline model.}
    \label{fig:interconnection_structure}
\end{figure}

We augment the baseline model \eqref{eq:FP_model} with learning components in terms of parameterized functions, jointly denoted by $\phi_\text{aug}$. The proposed LFR-based augmentation structure combines $\phi_\text{base}$ and $\phi_\text{aug}$ as shown in Fig. \ref{fig:interconnection_structure} and can be written as
\begin{subequations}\label{eq:interconnection}
     
    \begin{align}
        \hspace*{-2.9mm}  
        \left[\begin{array}{c}
        \hat x_{k+1} \\
        \hat y_k \\
        \hdashline[5pt/2pt]
        z_{1,k} \\
        z_{2,k} \\
        \end{array}\right] & \hspace*{-1mm} = \hspace*{-1mm}
        \underbrace{\left[\begin{array}{cc;{5pt/2pt}cc}
        \mathit{S_{x x}} & \mathit{S_{x u}} & \mathit{S_{x w_1}} & \mathit{S_{x w_2}}\\
        \mathit{S_{y x}} & \mathit{S_{y u}} & \mathit{S_{y w_1}} & \mathit{S_{y w_2}}\\
        \hdashline[5pt/2pt]
        \mathit{S_{z_1 x}} & \mathit{S_{z_1 u}} & \mathit{S_{z_1 w_1}} & \mathit{S_{z_1 w_2}} \\
        \mathit{S_{z_2 x}} & \mathit{S_{z_2 u}} & \mathit{S_{z_2 w_1}} & \mathit{S_{z_2 w_2}} \\
        \end{array}\right]}_\mathbf{S}
        \hspace{-2pt} \hspace*{-1mm}  \left[\begin{array}{c}
            \hat x_k \\
            u_k \\
            \hdashline[5pt/2pt]
            w_{1,k} \\
            w_{2,k} \\
        \end{array}\right] \\
        w_{1,k} & = \mathnormal{\phi}_\text{base}\left(z_{1,k}\right) = \begin{bmatrix}
            f_\text{base}\left(\theta_\text{base}, z_{1,k}\right)\\
            h_\text{base}\left(\theta_\text{base}, z_{1,k}\right)
        \end{bmatrix}\\
        w_{2,k} & = \mathnormal{\phi}_\text{aug}\left(z_{2,k}\right),
    \end{align}
\end{subequations}
where $\mathbf{S}\in \mathbb{R}^{n \times m}$ is the interconnection matrix and $\mathit{S}$ are selection matrices of appropriate size.

\subsection{Conditions on the interconnection matrix}
The interconnection matrix $\mathbf{S}$ can be fixed or parameterized. In this work, we consider fixed interconnection matrices $\mathbf{S}$. By further choosing~$\phi_\text{aug}$ formulations, the model augmentation structures in Table~\ref{tab:augmentation_structures} can be readily represented, as well as a wide variety of structures beyond them. The following condition must be considered for these formulations.
\begin{condition}[Well-posedness]
    For every state $\hat x_k$ and input $u_k$, model \eqref{eq:interconnection} admits a unique solution $z_k = \begin{bmatrix} z_{1,k} & z_{2,k} \end{bmatrix}$.
\end{condition}
We enforce this condition by avoiding algebraic loops in the formulations of \eqref{eq:interconnection}. This is equivalent to avoiding cycles in a graph as in the following theorem:
\begin{theorem}[Ayclic directed graph \cite{bang2008digraphs}]
    The~directed graph represented by the adjacency matrix $\mathbf{S}$ is acyclic if and only if it has a topological ordering.
\end{theorem}
The presence of a topological ordering can be computed in linear time\cite{bang2008digraphs}.

\subsection{Truncated loss function}
\begin{figure}[h]
    \centering
    \includegraphics[width=1.0\linewidth]{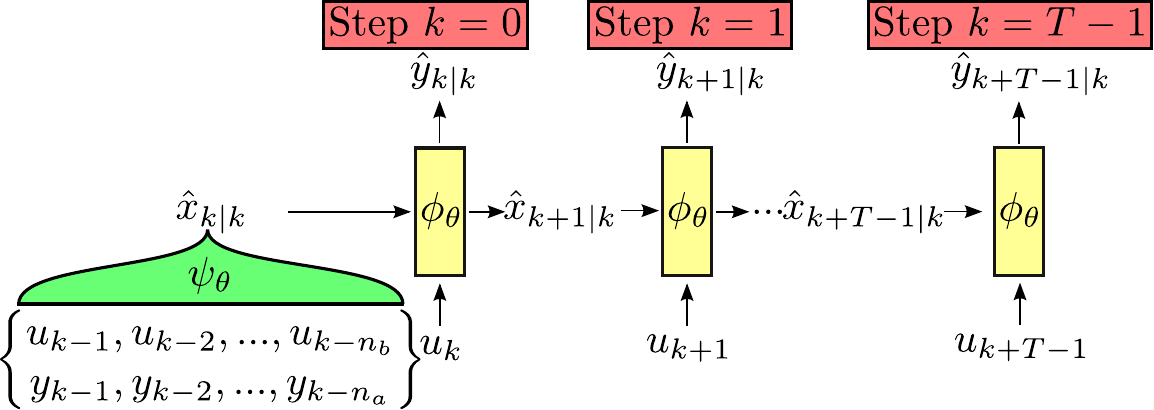}
    \vspace{2pt}
    \caption{SUBNET structure: the subspace encoder $\psi_{\theta}$ estimates the initial state at time index $k$ based on past inputs and outputs, then the state is propagated through  $\phi_\theta$ multiple times until a simulation length $T$.}
    \label{fig:n-step-encoder-graphic}
\end{figure}

In order to formulate the estimation of \eqref{eq:ss_model_structure}, the truncated objective function of SUBNET~\cite{gerben2022} is adapted. This allows for the use of computationally efficient batch optimization methods popular in machine learning, while also increasing data efficiency~\cite{gerben2022}. The truncated objective consists of $N$~subsections of length~$T$ of the available data~$\mathcal{D}_N$ as shown in Fig.~\ref{fig:n-step-encoder-graphic}. This objective function is given as
\begin{subequations} \label{eq:loss_function}
    \begin{align}
        \hspace*{-0mm} V_\text{trunc}(\theta) \hspace*{-0mm} =  \frac{1}{2 N(T+1)} & \sum_{i=1}^N \sum_{\ell=0}^{T-1}\left\|\hat{y}_{k_i \vert k_i+\ell}-y_{k_i+\ell}\right\|_2^2 \\
    \begin{bmatrix}
        \hat{x}_{k_i \vert k_i+\ell+1} \\
        \hat{y}_{k_i \vert k_i+\ell}
    \end{bmatrix} &:=\mathnormal{\phi}_\theta\left(\hat{x}_{k_i \vert k_i+\ell}, u_{k_i+\ell}\right) \\
    \hat{x}_{k \vert k} & :=\mathnormal{\psi}_{\theta}\left(y_{k-n_a}^{k-1}, u_{k-n_b}^{k-1}\right),
    \end{align}
\end{subequations}
where $\theta = \begin{bmatrix} \theta_\text{base}  & \theta_\text{aug} & \theta_\text{encoder}\end{bmatrix}$ is the joint parameter, $k \vert k + \ell$ indicates the state $\hat{x}_k$ or the output $\hat{y_k}$ at time $k + \ell$ simulated from the initial state $\hat{x}_{k \vert k}$ at time $k$; and $\phi_{\theta}$ is the LFR-based augmentation structure~\eqref{eq:interconnection} consisting of $\phi_\text{base}$, $\phi_\text{aug}$ and $\mathbf{S}$. The subsections start at a randomly selected time $k\in \{n+1,\ldots,N-T\}$. The initial state of these subsections is estimated by an encoder function $\psi_{\theta}$ from past input-output data, i.e., $\hat{x}_{k|k}=\psi_{\theta}(y_{k-n_a}^{k-1}, u_{k-n_b}^{k-1})$ where $u_{k-n_b}^{k-1}= [\ u_{k-n_b}^\top \ \cdots \ u_{k-1}^\top ]^\top$ for $\tau\geq 0$ and $y_{k-n_a}^{k-1}$ is defined similarly. This encoder is co-estimated with $\phi_{\theta}$.

\subsection{Regularized loss function}
The LFR-based model \eqref{eq:interconnection} is overparameterized. The joint identification of \eqref{eq:interconnection} can therefore result in learning components canceling out part of the baseline model. As a result, the baseline model parameters $\theta_\text{base}$ will deviate from the initialized values, possibly leading to a loss of interpretability of the baseline model. To address this issue, we adapt the regularization cost term from \cite{Bolderman2024}, which is given as
\begin{equation}
    V_\text{reg}(\theta) = \left\|\Lambda \left(\theta_\text{base} - \theta_\text{base}^*\right)\right\|_2^2,
\end{equation}
where $\Lambda$ is a matrix specifying the relative importance of each baseline model parameter. This matrix is computed as
\begin{equation}
    \Lambda = \left(\frac{1}{\epsilon}V_\text{MSE}(\theta_\text{base}^*)\right)^{\frac{1}{2}} \text{diag}(\theta_\text{base}^*)^{-1},
\end{equation}
where $V_\text{MSE}$ is the mean-square-error loss over for the baseline model against the training data, and $\epsilon$ is a tuning parameter specifying the permitted deviation from the initial parameter values. The final cost function with the regularization cost term becomes
\begin{equation}
    V(\theta) = V_\text{trunc} + V_\text{reg}.
\end{equation}
\subsection{Baseline model normalization}
For the estimation of ANNs, normalization of the input and output data to zero mean and to a standard deviation of~1 has shown to improve model estimation \cite{bishop1995neural}. Therefore, the LFR-based model augmentation structure is normalized as in \cite{gerben2022}, while the learning components are initialized as in \cite{Schoukens2020}. This means that $\hat x$, $u$, $\hat y$ in \eqref{eq:interconnection} are normalized. The to-be-augmented baseline model thus needs to have normalized input, state, and output as well. This is accomplished by the following transformation
\begin{subequations}
    \begin{align}
    \bar{f}_{\text{base}} & = T_x f_{\theta_\text{base}, \text{base}}\left(T_x ^{-1} \tilde x, T_u ^{-1} u \right) \\
    \bar{h}_{\text{base}} & = T_y h_{\theta_\text{base}, \text{base}}\left(T_x ^{-1} \tilde x, T_u ^{-1} u \right),
\end{align}
\end{subequations}
where $T_u \in \mathbb{R}^{n_u}$ is a diagonal matrix of the inverse standard deviations $\sigma_u^{-1}$. The standard deviations $\sigma_u$ are derived from $\mathcal{D}_N$. $T_y$ is similarly defined. For $T_x$, the standard deviation $\sigma_x$ is determined in the baseline states of the model $\tilde x$. These states can be obtained by, e.g., simulation of the baseline model with nominal input and initial conditions.

\subsection{Parameter estimation and properties}
The encoder-based model augmentation method can readily be implemented with PyTorch. In particular, the use of the ADAM optimizer\cite{Kingma2015} provides an efficient training method for the considered cost function and model structure. The original SUBNET method has been shown to provide consistency guarantees under the assumption of persistently exciting data and convergence of the optimization to the global minimum of the cost function~\cite{gerben2022}. Under the assumption that the system~\eqref{eq:nl_dyn} is part of the model set spanned by the parameterization of~\eqref{eq:interconnection}, the consistency properties of the SUBNET method are inherited.

\section{Simulation Results} \label{sec:Example}
\subsection{Mass-Spring-Damper system}
\begin{figure}[]
    \centering
    \includegraphics[width=0.49\textwidth]{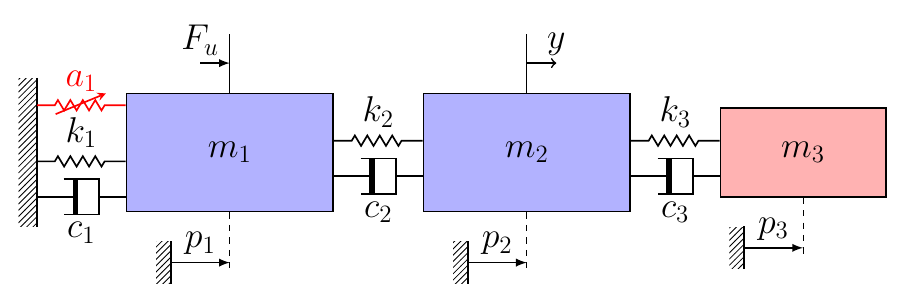}
    \caption{3-DOF MSD. The linear dynamics of the $m_1$ and $m_2$ are assumed known, while the dynamics of $m_3$ and the contribution of $a_1$ are unknown.}
    \label{fig:3dof_msd}
\end{figure}

As a simulation example\footnote{\scriptsize\texttt{https://github.com/Mixxxxx358/Model-Augmentation-Public}}, a 3 \emph{degrees-of-freedom} (DOF) \emph{mass-spring-damper} (MSD) is considered with a hardening spring nonlinearity, as shown in Fig. \ref{fig:3dof_msd} with the physical parameters as given in Table \ref{tab:params_msd}. The states associated with the system representation are the positions $p_i$ and the accelerations $\dot p_i$ of the masses $m_1$, $m_2$, $m_3$. The hardening spring is described as a cubic nonlinearity with parameter $a_1$. The position $p_2$ is the measured output.

\begin{table}[t]
    \centering
    \caption{Physical parameters 3-DOF MSD.}
    \vspace*{5pt}
    \label{tab:params_msd}
    \begin{tabular}{c||r|r|r|r} 
        Body & Mass $m_i$ & Spring $k_i$ & Damper $c_i$ & Hardening $a_i$ \\
        \hline
        1  & $0.5 \mathrm{~kg}$ & $100 \mathrm{~\frac{N}{m}}$ & $0.5 \mathrm{~\frac{N s}{m}}$  & $100 \mathrm{~\frac{N}{m^3}}$\\
        2 & $0.4 \mathrm{~kg}$  & $100 \mathrm{~\frac{N}{m}}$ & $0.5 \mathrm{~\frac{N s}{m}}$ & -\\
        3 & $0.1 \mathrm{~kg}$  & $100 \mathrm{~\frac{N}{m}}$ & $0.5 \mathrm{~\frac{N s}{m}}$ & -\\

\end{tabular}
\end{table}

\begin{table}[t]
    \centering
    \caption{Approximate physical parameters 2-DOF MSD baseline model.}
    \vspace*{5pt}
    \label{tab:approx_params_msd}
    \begin{tabular}{c||r|r|r} 
        Body & Mass $m_i$ & Spring $k_i$ & Damper $c_i$ \\
        \hline
        1  & $0.5 \mathrm{~kg}$ & $95 \mathrm{~\frac{N}{m}}$ & $0.45 \mathrm{~\frac{N s}{m}}$ \\
        2 & $0.4 \mathrm{~kg}$  & $95 \mathrm{~\frac{N}{m}}$ & $0.45 \mathrm{~\frac{N s}{m}}$ \\

\end{tabular}
\end{table}

We obtain the data for model estimation by applying 4th~order Runge-Kutta (RK4) based numerical integration on the 3-DOF MSD system with~$T_\mathrm{s}=0.02$~$\mathrm{s}$. The system is simulated for a zero-order hold input signal $F_u$ generated by a DT multisine $u_{k}$ with 1666 frequency components in the range $[0, 25]$ Hz with a uniformly distributed phase in $ [0, 2\pi)$. The sampled output measurements $y_k$ are perturbed by additive noise $e_k$ described by a DT white noise process with $e_k\sim\mathcal{N}(0,\sigma_\mathrm{e}^2)$. Here, $\sigma_\text{e}$ is chosen so that the signal-to-noise ratio equals 30 dB. After removing the transient due to initial conditions, the sampled generated output $y_k$ for the input~$u_{k}$ is collected in the data sequence $\mathcal{D}_N$. A separate data sequence is generated for estimation, validation, and testing with size $N_{\text{est}} = 2\cdot10^4$, $N_{\text{val}} = 10^4$, $N_{\text{test}} = 10^4$, respectively.

\subsection{Baseline model}
The baseline model is chosen to represent the linear 2-DOF MSD dynamics corresponding to the blue masses in Fig.~\ref{fig:3dof_msd}, i.e, both the red mass $m_3$ and the nonlinear spring $a_1$ are left out. We consider two initializations for the baseline model parameters: the ideal values from Table~\ref{tab:params_msd}, and approximate values from Table~\ref{tab:approx_params_msd}. The \emph{root mean squared simulation error}~(RMSE) of the baseline model for both initializations are shown in Table~\ref{tab:nrms}. The simulation error of the baseline model with ideal values on the test data is shown (orange) in Fig.~\ref{fig:simulation_error}.

\begin{table}[t]
    \centering
    \caption{Root mean squared simulation error of the baseline and estimated models.}
    \begin{tabular}{c||c|c}
    Model \!&\! Ideal parameters \!&\! Approximate parameters \\
    \hline
    baseline & 0.0583 & 0.0795 \\ 
    \hline
    static series & 0.00798 & 0.00827 \\
    dynamic series & 0.00601 & 0.00552 \\
    static parallel & 0.00563 & 0.00567 \\
    linear dynamic parallel & 0.00972 & 0.0113 \\
    dynamic parallel & \bf{0.00534} & \bf{0.00529} \\
    \hline
    ANN-SS & \multicolumn{2}{c}{0.00557}\\
    \end{tabular}
    \label{tab:nrms}
\end{table}

\subsection{Estimated models}
We estimate variations of the LFR-based augmentation structure corresponding to representations of the augmentation structures described in Table~\ref{tab:augmentation_structures}. For parallel augmentation structures, the learning components are chosen as feedforward neural networks, and for series augmentation structures, we choose residual networks (ResNet)\cite{he2016deep}. ResNets are chosen to capture the dominant linear behavior of the system in the series augmentation structure. This is not necessary for the parallel model, as the baseline model is additively augmented. For all learning components, the number of hidden layers and neurons are as in Table~\ref{tab:Hyperparam}. The activation function is chosen as $\tanh$ for all models except the linear dynamic parallel model, which uses an identity activation function to represent a linear augmentation. For the dynamic augmentation, we add two additional states to the baseline model states for a total of 6 states. This is the minimum number of states required to model the 3-DOF MSD system. The baseline model, learning component, and encoder parameters are jointly estimated as described in Section~\ref{sec:Method}. The hyperparameters for these estimations are shown in Table~\ref{tab:Hyperparam}. The regularization tuning parameter $\epsilon$ in the joint identification cost function, was determined through a line search. The optimal value was determined at $\epsilon=1$. As a comparison to a black-box approach, an ANN-SS model parameterized by ResNets is estimated with the same estimation procedure and hyperparameters.

\begin{figure}[]
    \centering
    \includegraphics[width=\linewidth,trim={0 0 0 35pt},clip]{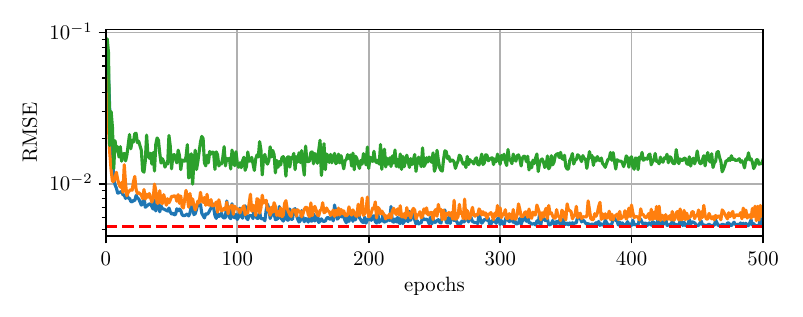}
    \caption{Validation loss over first 500 training epochs for dynamic parallel (blue), static parallel (orange) and linear dynamic parallel (green) models. The noise is floor is shown with a dashed line.}
    \label{fig:val_loss}
\end{figure}

The simulation RMSE on the test data for these estimated models are shown in Table~\ref{tab:nrms}. The parallel augmentation models reach a steady validation loss after 500 epochs as shown in Fig.~\ref{fig:val_loss}. Among the parallel augmentation models, the dynamic performs best, followed by the static and linear dynamic augmentation models, respectively. This indicates that the inclusion of additional states and nonlinear augmentations is beneficial for obtaining accurate augmentation models of the 3-DOF MSD system for the given baseline model.

The dynamic series and ANN-SS models achieve RMSE scores similar to those of dynamic parallel augmentation as shown in Table~\ref{tab:nrms}. They, however, require up to 5000 epochs to reach a steady validation loss.

For all augmentation models, the final estimated physical parameters are close to the initialized values, both the exact and approximate initializations. For the approximate initialization, this indicates that the learning components are learning part of the system dynamics that could be represented by the baseline model. 

\begin{table}[t]
    \centering
    \caption{Hyperparameters for identifying the LFR-based augmentation models and ANN-SS model.}
    \vspace*{4pt}
    \begin{tabular}{c|c|c|c|c|c|c}
    \!hidden layers\!&\!nodes\!&\!$n_u$ $n_y$\!&\!$n_a$ $n_b$\!&\!$T$\!&\!epochs\!&\!batch size\!\\
    \hline
    2 & 64 & 1 & 7 & 200 & 5000 & 2000 \\
    \end{tabular}
    \label{tab:Hyperparam}
\end{table}

\begin{figure}[]
    \centering
    \includegraphics[width=\linewidth]{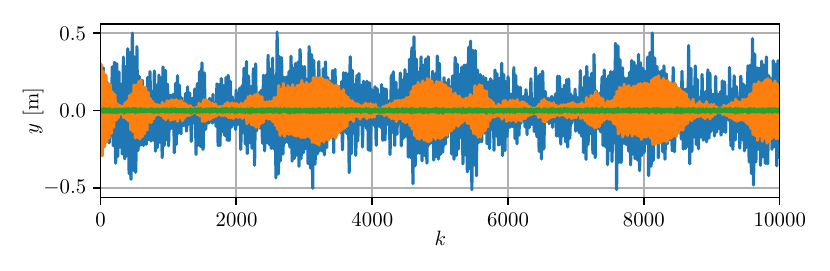}
    \caption{Simulation error of the baseline model (orange) and nonlinear state augmented model (green) compared to the measured test data (blue) for the MSD system.}
    \label{fig:simulation_error}
\end{figure}

\begin{figure}[]
    \centering
    \includegraphics[width=\linewidth,trim={0 0 0 35pt},clip]{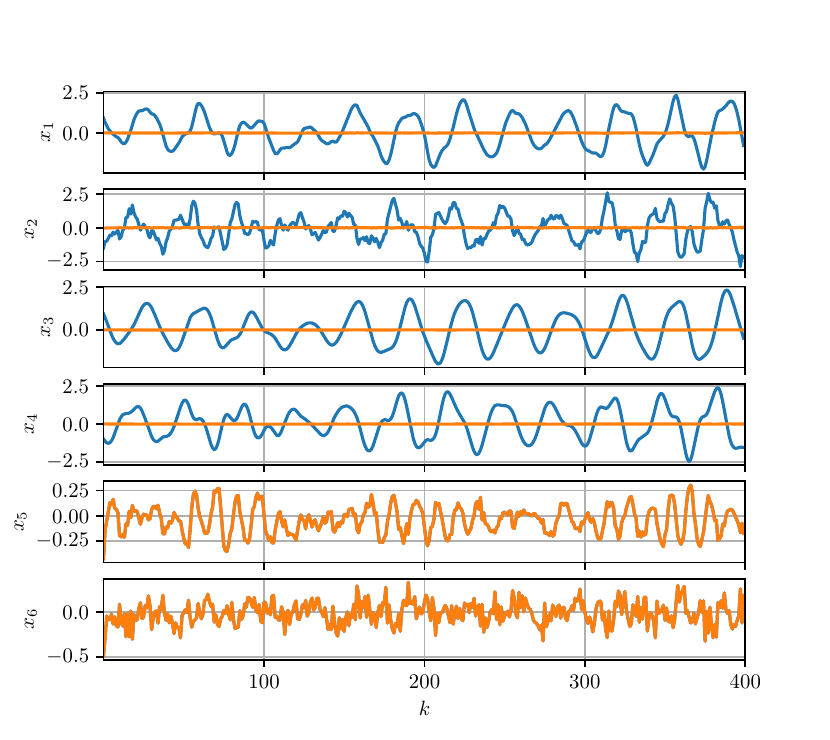}
    \caption{Comparison between dynamic parallel model states $\hat x$ (blue) and the outputs of the learning components $\phi_\text{aug}$ (orange) for a simulation with test data. States $x_1$ - $x_4$ are the sum of the baseline model $f_\text{base}$ and the parallel augmentation $f_\text{aug}$, while states  $x_5$ - $x_6$ are the output of the dynamic augmentation $g_\text{aug}$.}
    \label{fig:state_comparison}
\end{figure}

Further analysis of the dynamic parallel model is provided in Fig.~\ref{fig:simulation_error} and Fig.~\ref{fig:state_comparison}. In Fig.~\ref{fig:simulation_error}, we show the simulation error on test data (blue) of the nonlinear dynamic parallel augmented model (green) and the baseline model (orange). The figure shows consistent error performance across all test data. In Fig.~\ref{fig:state_comparison}, we show the comparison between the states~$\hat x$ of the dynamic parallel model (blue) and the outputs of the learning components $\phi_\text{aug}$ (orange). Here, $\tilde x = \begin{bmatrix} x_1 & \ldots & x_4 \end{bmatrix}$ and $\bar x = \begin{bmatrix} x_5 & x_6 \end{bmatrix}$. The output of the learning components is relatively small for $\tilde x$, and $\bar x$ is modeled solely by the learning components. From this we can conclude that the learning components are augmenting the baseline model resulting in an accurate model and not replacing the baseline model with its own dynamics. Therefore, we have obtained an accurate model whose dominant behavior is given by an interpretable prior model.

\section{Conclusion} \label{sec:Conclusion}
We have adapted the encoder-based estimation methods for the model augmentation setting. For this, we have introduced a flexible LFR-based model augmentation structure. In an MSD simulation, we have shown how the proposed estimation method and LFR-based structure can be used for static and dynamic model augmentation. The resulting models are accurate, estimated efficiently, and retain the interpretability of the baseline model. We also observe that the baseline model parameters remain close to the initializations, preserving interpretability.

Fixed interconnections are considered in this work. Future research will investigate parameterized interconnections for use in automatic model structure selection.

\bibliographystyle{IEEEtran}
\bibliography{references}

\end{document}